\magnification 1200
\nopagenumbers
\parskip=10pt
\vskip .7truein
\centerline {\bf{ Gravitational shock waves and vacuum fluctuations}}
\vskip 1.0truein
\centerline { M.Horta\c csu and K. \" Ulker }
\vskip .3truein
\centerline { Physics Department, Faculty of Sciences and Letters}
\vskip .1truein
\centerline {I.T.U., 80626 Maslak, Istanbul, Turkey}
\vskip 1.4truein
Abstract: We show that the vacuum expectation value of the 
stress-energy tensor of 
a scalar particle on the background of a spherical gravitational shock wave 
does not give a finite expression in second order perturbation theory, 
contrary to the case seen for the impulsive wave.   No infrared divergences
appear at this order.  This result shows that there is a qualitative
difference between the shock and impulsive wave solutions which is not
 exhibited in first order.
\vfill\eject
\pageno=1
\noindent INTRODUCTION

Both from physical and mathematical points, the cosmic string solutions
$^{/1}$ of Einstein's field equations are interesting $^{/2}$.
 An immediate question is whether these strings decay.
  Exact solutions describing such decays are given
for  impulsive waves by Nutku-Penrose $^{/3}$ and Gleiser and Pullin
$^{/4}$ , for  shock waves by Nutku $^{/5}$.

Once these solutions are found one may question whether they give rise to
vacuum fluctuations.  We have investigated these fluctuations in several
papers.  In first order perturbation theory, we found that we could not
isolate a finite part for the vacuum expectation value , VEV, of
the stress-energy
tensor both for the impulsive $^{/6}$ and shock wave solutions $^{/7}$.
When the calculation is carried to second order for the impulsive wave case,
a finite result is found $^{/8}$ if a detour is taken to  de Sitter space.
The essential point in this calculation is the generation of an infrared
 divergence
in second order perturbation theory which is regulated by an infrared mass.
We  go to de Sitter space and there cancel this mass by the cosmological
constant.  At the end we let both the infrared mass and the cosmological
constant go to zero and  obtain a finite result.

In this note we carry the calculation  in the background 
of the shock wave metric   to second order and
 investigate whether the same trick gives us a finite
expression for the VEV of the stress-energy tensor for this case 
 .  If the calculation does not generate an infrared divergence ,
going to de Sitter space gives us a finite result only in this space
which vanishes when we go back to Minkowski space $^{/9}$.

 In first order perturbation theory both the impulsive
and the shock wave cases showed similar behaviour.  The two solutions are
essentially different, though.  The shock wave solution has a dimensional
constant which is lacking in the impulsive wave solution.  Since
 in quantum field theory , models
 with dimensional and dimensionless constants belong to different classes,
  we thought similar distinction
 between these two models may exist. Relying on these motivations we planned
 to check whether there is a qualitative difference 
between the two solutions exhibited by their behaviour at higher orders .

We will show that the infrared divergences which may be cancelled via a 
detour in de Sitter space are  absent in the shock wave calculation.
Another point of difference is the importance attributed to the homogenous
solutions
in these two cases.
 The homogenous solutions give just the free Greens function for the
impulsive case, whereas they result in a totally
different contribution to the
Greens function in the shock wave.
This is an artefact of the presence of a dimensional coupling
constant in the latter case.
 What may be more interesting is the fact that
just the contribution of the first order calculation
 contributes to $<T_{\mu \nu}>$ in de Sitter space.
The higher order terms cancel out when the VEV of the stress-energy tensor is
computed.

We give our calculation in the next section. We calculate the Greens function and the VEV 
of the stress-energy tensor exactly as we have done in references 6-9.  These
methods are more thoroughly described in $^{/10}$, details are in  
   $^{/11}$.
We conclude with few remarks.
\vfill\eject
\noindent CALCULATION

We start with the metric
$$ ds^2= 2P du dv + 2u P_{\zeta} d\zeta dv + 
2u P_{\overline{\zeta}} d{\overline{\zeta}}dv
-2u^2 d\zeta d\overline{\zeta} \eqno {1} $$
whose properties are described in references 5 and 7.  Here 
$P={1\over {|h_{\zeta}|}}$,
where $h$ is an arbitrary function of the argument $\zeta+gv \Theta(v)$.
 $g$ is the dimensional coupling constant and $\Theta$ is the Heavyside 
unit step function.
In our particular case we take $h=(\zeta+gv\Theta(v))^{1+i\delta}$, where
$\delta<<1$ and is the expansion parameter.

We expand the operator $L$ which is equal to $\sqrt {-g} $ times the
d`Alembertian operator, in powers of $\delta$.
$$L=L_0+\delta L_1 + {\delta}^2 L_2+... $$
If we use real variables, $\zeta ={{x+iy}\over {\sqrt 2}}$, we get
$$ L_0= 2u^2{{\partial ^2}\over { \partial u \partial v}}+
2u{{\partial}\over {\partial v}}-{{\partial^2} \over {\partial x^2}}
-{{\partial^2} \over {\partial y^2}}, 
\eqno {3} $$
$$L_1= -{{2u}\over {x^2+y^2}} \left( y {{\partial}\over {\partial x}}
-x{{\partial} \over {\partial y}} \right) {{\partial } \over {\partial u}}
-tan^{-1}{y \over {x}} \left( {{\partial^2} \over {\partial x^2}}
+{{\partial^2} \over {\partial y^2}} \right) ,\eqno {4} $$
$$L_2=-{1 \over {x^2+y^2}} \left( u{{\partial} \over {\partial u}}+
u^2 {{\partial^2}\over {\partial u^2}}+ 2 tan^{-1} {y \over {x} }
\left(y{{\partial} \over {\partial x}} -
x{{\partial} \over {\partial y}} \right)
u {{\partial} \over {\partial u }} \right) $$
$$+{1\over {2} } \left( 1-(tan^{-1} {y \over {x}})^2 \right) 
\left( {{\partial ^2}\over {\partial x^2}}+{{\partial ^2}\over
{\partial y^2}}
\right) . \eqno {5} $$
We expand both the solutions and the eigenvalue in terms of $\delta$,
$\phi=\phi_0+\delta \phi_1+\delta^2 \phi_2+...$,$\lambda=\lambda_0+
\delta \lambda_1
+\delta^2 \lambda_2+... $ .  We take $\phi_1=f \phi_0$, $\phi_2=h\phi_0.$
Here $f=f_0(z,y,u)+gf_1(z,y,u), h=h_0(z,y,u)+gh_1(z,y,u)$ where
 $z=x+gv\Theta(v)$.
$\phi_0$ is given by 
$$\phi_0= {{\exp[i(k_1x+k_2y+Rv-{K\over {2Ru}})]}\over 
{u\sqrt{|R|} (2\pi)^2}}. \eqno {6} $$   $K,k_1,k_2 ,R$ are the
seperation constants which act as eigenfrequencies to be integrated
 over to find the Greens Function.

In second order in $\delta$, we can reduce the differential equation to
 the system
$${\cal L}_0 h_0 = I_0 ,\eqno {7}  $$
$${\cal L}_0 h_1 +{\cal L}_1 h_0 = I_1 , \eqno {8} $$
where
$${\cal L}_0 =-2iR {{\partial} \over {\partial s}} -2i \left( k_1
{{\partial}\over {\partial z}}
+k_2 {{\partial} \over {\partial y}} \right)
-{{\partial ^2} \over{\partial z^2}}-
{{\partial^2} \over {\partial y^2}}, \eqno {9} $$
$${\cal L}_1 =-2{{\partial ^2}\over {\partial s \partial z}} +{iK \over {R} }
{{\partial}\over {\partial z}}.\eqno {10} $$
Here $s={1\over{u}}$.  $I_0$ and $I_1$ are given as
$$I_0={1\over{2}}K-{1 \over {z^2+y^2}}
\left( 1-{3iKs \over {2R}}-{K^2s^2 \over {4R^2}} \right)
\left(1-2i (k_1y-k_2z) tan^{-1} {y \over {z}} \right)$$
$$+K\left( {5 \over {2}}-{iKs \over {2R}} \right)(tan^{-1} {y\over {z}})^2$$
$$-2K \left( \left({log (z^2+y^2) \over {2}}-1 \right)(k_1y-k_2z)
+tan^{-1}{y\over {z}}(k_1z+k_2y) \right) $$
$$ \times \left(\left(1-{iKs \over {R}}\right)
  {{k_1y-k_2z} \over {(z^2+y^2) K}}
-{i\over {2}} tan^{-1} {y\over {z}} \right)             $$
$$-{i \over {z^2+y^2} } \left( (1-{iKs \over {2R}}) 
\left( (k_1z+k_2y) log (z^2+y^2) +2(k_2z-k_1y) tan^{-1}
{y \over {z}} \right) \right)
\eqno {11} $$
$$I_1={{2iyK}
\over{(z^2+y^2)R}}\left(1-{{iKs}\over {4R}} \right) tan^{-1} {y \over z}
-{2k_1 \over {(z^2+y^2)R}}\left(1-{iKs \over {R}}-
{K^2s^2 \over {8R^2}}\right)
$$
$$- \left( 2k_1 tan^{-1} {y\over z} -k_2 log(z^2+y^2) \right)$$        
$$\left( \left({{-iK^2 s}\over {8R^2}}+{{5K}\over {4R}} \right)
  tan ^{-1} {y \over z} - {{i(k_1y-k_2z)} \over {(z^2+y^2)R}}
\left( 1-{{iKs}\over {R}}-{{K^2 s^2} \over {8R^2}} \right) \right)
    $$
$$+\left((k_1^2-k_2^2) \left( 2z tan^{-1}{y\over z}+
y log(z^2+y^2)-2y \right)
-2k_1k_2 \left( z log(z^2+y^2)
 -2y tan^{-1} {y \over z} -2z \right) \right) $$
$$ \times \left( {3Ki \over {8R}} tan ^{-1} {y \over z} 
-{3 \over {4(z^2+y^2)R}}(k_1y-k_2z)(1-{iKs \over {2R}}) \right)        
-{3i \over {4(z^2+y^2)R}}  
  \left(1-{iKs \over {2R}} \right)$$ 
$$\times \left( (k_1^2-k_2^2)
\left(z log(z^2+y^2) -2y tan ^{-1} {y \over z} \right)
+2k_1k_2 \left( y log(z^2+y^2)+2z tan ^{-1} {y \over z} \right) \right)
\eqno {12} $$
In these expressions we took the `mass-shell' condition,
 which is imposed in the
calculation of the Greens function; i.e. we set $k_1^2+k_2^2$ equal to $K$.
  One can
check that after we perform the $ K$ and $k_1, k_2$ integrations
the effect of these two expressions are exactly the same.

We see that, contrary to the impulsive wave calculation,
in both $I_0$ and $I_1$, there are no terms that are independent
of $z$ and $y$ except a single term which is proportional to $K $.
To be able to obtain terms in $<T_{\mu \nu}>$
that diverge as the infrared parameter goes to zero, we need inverse powers
 of $R$ which are not multiplied by $K$ or $k_1^2, k_2^2$.
    Each inverse power of $R$ means a higher infrared divergence,
     order going as $m^2,1, \log m^2,{1\over {m^2}},{1\over {m^4}}$, etc...,
whereas each power of $ K,k_1^2,k_2^2$ means one lower order in the same
 divergence. In free space the power of $m$ is zero.
   There is no  divergence.

In reference 4, we generated these divergences at second order and then
 cancelled them with the cosmological constant of the de Sitter solution.
 Our mechanism for obtaining these infrared divergences was as follows.
  We isolated the $ -2iR {{\partial} \over { \partial s}}$
in the operator ${\cal L}_0 $ from the others and equated it to the
term which did not contain $ z $ or $y$.
$$-2iR {{\partial} \over {\partial s}} h'_0 =cs  \eqno {13}$$
where $c$ can be a function of $v$ but not that of $z$ and $y$.  Then 
$h'_0 = {{ics^2}\over { 4R}}$ which has an extra power of $1/R$
 compared to the other terms.
The second iteration gives us $h'_1 \propto s^3/{R^2} $.
Such a term will induce $1/m^2$ factor in the expression for
the Greens Function , $G_F$, and this infrared
mass will be retained in $<T_{\mu \nu}>$.

For the shock wave solution all the terms in $I_0$ and $I_1$ are either
functions of $z$ and $y$ or are multiplied by $K$.  We can not isolate a part
 of the operator ${\cal L}_0 $ and equate it to a single term on the RHS.
  Note that in the previous argument we may act in this way since the
  rest of the operators in ${\cal L}_0$
will annihilate the resulting expression $h'_0$.
 If this terms is not annihilated by the other operators, there will be a
 mismatch in the powers in $s$ on both
sides of the equation.  To illustrate this let's assume we had
$$-2iR{{\partial}\over {\partial s}}h'_0= f(z,y)s^2. \eqno {14}$$ 
Upon integration we get 
$h'_0 ={if(z,y) s^3 \over {6R}}$.  Such a term will not be annihilated
 when rest of the terms in ${\cal L} _0 $ operate on it and we will
 generate a higher power of $s$ than  the one
we have started from, which has no match on RHS.  Since $I_0$ and $I_1$
do not contain powers of $s$ higher than the quadratic,
 there is no way we can generate terms with the third power of $s$ in this
  way, also no way to generate ${1\over {R^3}}$ which will multiply such
   a term.  Similarly we can show that we can not generate
a power of ${1\over{R}}$  in a combination which does not already exist
on RHS.  On RHS
only the combination ${K^2 \over {R^2}} $ and ${K \over {R^2}} $ exist .
  ${K^2 \over {R^2}}$ gives exactly
the singularity structure as the free case, and ${K \over {R^2}}$ gives a
logarithmic divergence
which is cancelled in the $<T_{\mu \nu}>$ calculation.

At this point we note that we can find solutions of equations 9 and 10
  even if
$I_0$ and $I_1$ are set to zero.  These are the homogenous solutions of the
problem which give a non trivial contribution for the shock wave calculation.
Since $I_0$ and $ I_1$ are independent of $v$ we can assume a powers series
expansion in $v$ for a chosen order in $g$.  For the sake of illustration we
take a solution in third order in $g$ and write the expansion as
$$ f^{(1,3)}_H = g^3 ( v^3 f^{(1,3)}_{1H} + v^2 f^{(1,3)}_{2H}
+v f^{(1,3)}_{3H}+f^{(1,3)}_{4H}). \eqno {15} $$
Here $f_{1H}^{(1,3)}(s,z,y)$ has dimension zero, $f_{2H}^{(1,3)}(s,z,y)$
 has dimension
 minus one, etc.. Inverse powers of $v$ are excluded by the regularity
 at $v=0$.  One can show that taking powers of
$v$ higher than that of $g$ do not give  results that differ from the
 free case.  A similar expansion in the impulsive case would go as
 $ f_{H}^{(1,3)}=(v^3R^3 f_{1H}^{(1,3)}+v^2R^2 f_{2H}^{(1,3)} +...)$
 when $f_{1H}^{(1,3)}$ etc., have the same dimensions as above,
 since the only
 free dimensional parameters are $v$ and $R$.  This gives the free result.

Keeping track of
powers of $v$ we get a system of four equations.  We note that the first of
these equations 
$${\cal L}_0 f^{(1,3)}_{1H} = 0 \eqno {16} $$
has a solution for any function $F=F({s \over {R}}(k_1 \pm i k_2)
-(z \pm iy))$.  We can also show that the singularity behaviour of the
Greens Function is independent of the form of $F$.  At the end we get,
 for the worse infrared poles the expressions
$$ G_H^{(1,H)} =2 \pi c_1 \left( {{v^3 \Theta(v) + v'^3 \Theta(v')}\over
 {(u-u')(v-v')}} \right) ,\eqno {17} $$
$$ G_H^{(2,H)} = 2 \pi c_2 \left( {{ uv^2 \Theta(v) + u'v'^2 \Theta(v')}
 \over {(u-u')^2}} \log (2m^2(u-u')(v-v')) \right), \eqno {18} $$
$$ G_H^{(3,H)} = 2 \pi c_3 \left( {{ u^2v \Theta(v)+u'^2v' \Theta(v')}
\over {(u-u')^4 m^2}} \right)   , \eqno {19} $$
$$ G_H^{(4,H)} = 2 \pi c_4 \left( {{ u^3 \Theta (v)+u'^3 \Theta (v')} 
\over {(u-u')^6 m^4}} \right). \eqno {20} $$
Here $c_i$ are functions of $x$ and $ y$, depending on the form for $F$ used.
$m$ is the infrared mass.
If we use linear function, then $c_i$ is proportional to $y$ or $x$.

Upon symmetric differentiation the terms with ${1 \over {m^2}}$ and
 $ {1 \over {m^4}} $ vanish.  We may find a finite contribution only if we
  go to the  de Sitter space, i.e. multiply by the factor
  $(1+{\Lambda uv \over {6}})(1+{\Lambda u'v' \over {6}}) $.
In this case we get
$$ <T_{vv}> = g^3 \left(  f_1 \Lambda v \Theta(v) +
f_2 \Lambda^2 uv^2 \Theta(v) \right) \eqno {21} $$
which goes to zero with $\Lambda$ when we go back to Minkowski background.  
In this expression $f_1$ are regular functions of $x,y$.
One can also show that any terms with less diverging powers of $m^{-2}$ 
in $G_H^{(4,H)}$ do not give a finite contribution even in
 de Sitter background.
\vskip .1truein
\noindent CONCLUSION

Here we tried to show that two qualitative differences exist
between the shock and the
impulsive wave solutions proposed by the same group $^{/3,5}$.
  In the shock wave
solution  the infrared divergences which may be used to tame the
 ultraviolet divergences
to  result in finite contributions to $<T_{vv}>$ are absent in second order
perturbation theory.  We can not find finite contributions
 to $<T_{v v}>$ in Minkowski
space.  If we go to de Sitter space, though, we get a finite contribution 
which is proportional to $\Theta$ function,
 which is the signature of a shock wave solution.

The homogenous solutions, in the shock wave, 
give contributions to the Greens function expression which are different
from the free case.  These solutions also give a finite contribution to 
$<T_{vv}>$ in de Sitter space.  The presence of these nontrivial solutions
is only due to the dimensional coupling constant.  The presence of $g$
in the expansion makes it necessary to have an extra power of ${1\over {R}}$
in the solution which results in a nontrivial term in $G_F$.

Ackowledgement:  We thank Prof.Dr.Y. Nutku for his support.  This work is
partially supported by TUBITAK, the Scientific and Technical Research
Council of Turkey and TUBA, the Turkish Academy of Sciences.

\vfill\eject
REFERENCES 

\item {1.}  J.R. Gott III, Astrophys. J. {\bf {288}} (1985) 442,
            W.A. Hiscock, Phys. Rev.D {\bf {31}} (1985) 3288,
            B. Linet, Gen. Rel. Grav. {\bf {17}} (1985) 1109;

\item {2.}  M.B. Hindmarsh and T.W.B. Kibble,
Reports Progress Phys. {\bf {58}} (1995) 471,
A.Vi\-len\-kin and E.P.S. Shellard, {\it{ Cosmic Strings and
other Topological Defects}}, Cambridge
University Press, Cambridge 1994,
 G.W.Gibbons, S.W. Hawking and T. Vachaspati (ed.), {\it{ The Formation
 and Evolution of Cosmic Strings}}, Cambridge University Press,
  Cambridge (1990);

\item {3.}  Y. Nutku and R. Penrose, Twistor Newsletter {\bf{ 34}} (1992) 9;

\item {4.}  R.Gleiser and J. Pullin, Class. Quantum Grav.
 {\bf{6}} (1989) L141;

\item {5.}  Y. Nutku, Phys. Rev. D {\bf{44}} (1991) 3164;

\item {6.}  M. Horta\c csu , Class. Quant. Grav. {\bf{7}} (1990) L165,
{\bf{9}} (1992) 799 (erratum),
J. Math. Phys. {\bf{34}} (1993) 690, M. Horta\c csu and R. Kaya,
 J. Math. Phys. {\bf{34}} (1990) 3043;

\item {7.} M. Horta\c csu, Class. Quantum Grav. {\bf{ 9}} (1992) 1619;
  N. \" Ozdemir and M. Horta\c csu, Class. Quantum Grav.
  {\bf{12}} (1995) 1221;

\item {8.}  M. Horta\c csu, Class. Quantum Grav. {\bf{13}} (1996) 2683;

\item {9.}  M.Horta\c csu and N.\" Ozdemir ITU Report (1995) .

\item {10.} N.D.Birrell and P.C.W. Davies, {\it{ Quantum Fields in
Curved Space}}, Cambridge University Press, Cambridge 1982;

\item {11.} K. \" Ulker, Thesis, submitted to Istanbul Technical
University 1997.

\end